# Orbital-selective band modifications in a charge-ordered kagome metal LuNb$_6$Sn$_6$

Rui Lou[1,2,3,*,†], Yumeng Zhang[4,5,*], Erjian Cheng[6,*,†], Xiaolong Feng[6,†], Alexander Fedorov[1,2,3], Zongkai Li[4], Yixuan Luo[4], Alexander Generalov[7], Haiyang Ma[8], Quanxing Wei[9], Yi Zhou[9], Susmita Changdar[1], Walter Schnelle[6], Dong Chen[9], Yulin Chen[4,10], Jianpeng Liu[4], Yanfeng Guo[4], Sergey Borisenko[1], Denis V. Vyalikh[11,12], Claudia Felser[6], Bernd Büchner[1,13], Zhongkai Liu[4,†]

[1] *Leibniz Institute for Solid State and Materials Research, IFW Dresden, 01069 Dresden, Germany*
[2] *Helmholtz-Zentrum Berlin für Materialien und Energie, Albert-Einstein-Straße 15, 12489 Berlin, Germany*
[3] *Joint Laboratory "Functional Quantum Materials" at BESSY II, 12489 Berlin, Germany*
[4] *School of Physical Science and Technology, ShanghaiTech Laboratory for Topological Physics, ShanghaiTech University, Shanghai 201210, China*
[5] *Zhangjiang Laboratory, 100 Haike Road, Pudong, Shanghai 201210, China*
[6] *Max Planck Institute for Chemical Physics of Solids, 01187 Dresden, Germany*
[7] *MAX IV Laboratory, Lund University, Fotongatan 2, 224 84 Lund, Sweden*
[8] *Quantum Science Center of Guangdong-Hong Kong-Macao Greater Bay Area, Shenzhen 518045, China*
[9] *College of Physics, Qingdao University, Qingdao 266071, China*
[10] *Department of Physics, University of Oxford, Oxford OX1 3PU, UK*
[11] *Donostia International Physics Center (DIPC), 20018 Donostia-San Sebastián, Spain*
[12] *IKERBASQUE, Basque Foundation for Science, 48011 Bilbao, Spain*
[13] *Institute of Solid State and Materials Physics, TU Dresden, 01062 Dresden, Germany*

*These authors contributed equally to this work
†Corresponding authors: lourui09@gmail.com, Erjian.Cheng@cpfs.mpg.de, Xiaolong.Feng@cpfs.mpg.de, liuzhk@shanghaitech.edu.cn

**The origin of the charge order in kagome lattice materials has attracted great interest due to the unique electronic structure features connected to kagome networks and the interplay between electron and lattice degrees of freedom. Recently, compounds with composition *Ln*Nb$_6$Sn$_6$ (*Ln* = Ce-Nd, Sm, Gd-Tm, Lu, Y) appear as a new family of kagome metals, structurally analogous to *R*V$_6$Sn$_6$ (*R* = Sc, Y, or rare earth) systems. Among them, LuNb$_6$Sn$_6$ emerges as a novel material hosting charge density wave (CDW) with a √3 × √3 × 3 wave vector, akin to that in ScV$_6$Sn$_6$. Here, we employ high-resolution angle-resolved photoemission spectroscopy, scanning tunneling microscopy, and density functional theory calculations to systematically investigate the electronic properties of LuNb$_6$Sn$_6$. Our observation reveals the characteristic band structures of the "166" kagome system. A charge instability driven by Fermi surface nesting is decisively ruled out through an analysis of the interactions between van Hove singularities. Across the CDW transition, we observe orbital-selective band modifications, with noticeable evolutions of Lu 5*d* and Sn 5*p* electrons, while Nb 4*d* electrons exhibit minimal change, suggesting that the Lu and Sn sites other than the Nb kagome lattice play a key role in the formation of CDW. Our findings substantiate a universal lattice-driven CDW mechanism rather than a charge-instability-driven one in the "166" kagome compounds, making it a distinct material class compared to other charge-ordered kagome systems, such as *A*V$_3$Sb$_5$ (*A* = K, Rb, Cs) and FeGe.**

## Introduction

The kagome lattice, made up of corner-sharing triangles, has emerged as a versatile playground for various emergent quantum phases. Due to the unique geometry of kagome lattice, its electronic band structure is characterized by three typical features, including the Dirac point (DP) at Brillouin zone (BZ) corner, the van Hove singularities (vHSs) at BZ boundary, and the flat band over the entire BZ[1–8]. These kagome band structures have been proposed to be responsible for a variety of novel quantum phenomena and symmetry-breaking states, such as the giant anomalous Hall effect[9], charge density wave (CDW)[10–18], nematic order[19–23], superconductivity[24–27], and pair density wave[28].

Recently, there has been considerable interest in the time-reversal symmetry-breaking CDW orders in kagome superconductors *A*V$_3$Sb$_5$ (*A* = K, Rb, Cs)[10–18] and kagome antiferromagnet FeGe[29–32]. The formation of 2 × 2 × 2 charge ordering could be intimately associated with the electronic nesting between vHSs near the Fermi level ($E_F$)[10,11,29,32,33]. The discovery of the intermetallic kagome compound ScV$_6$Sn$_6$ has offered an alternative platform for exploring novel CDW states[34]. In contrast to *A*V$_3$Sb$_5$ and FeGe, the CDW in ScV$_6$Sn$_6$ features a (√3 × √3) *R*30° × 3 reconstruction[34–40]. The kagome band structure has been suggested to play a minor role in triggering the CDW transition therein[41,42]; instead, the structural distortions of Sc and Sn atoms involving the electron-phonon coupling (EPC) could be the driving force[42–45]. However, the presence of a nematic-like intermediate phase within the CDW state[46,47] and a dynamic short-range CDW competing with the long-range

CDW[48–50] has been reported, rendering the origin of the CDW in ScV$_6$Sn$_6$ more intricate. Therefore, comparative studies on a parallel charge-ordered kagome system, which could shed light on the CDW mechanism, are highly desirable. In this regard, the newly discovered LuNb$_6$Sn$_6$[51], which shares the same CDW wave vector and similar primary distortions along the Lu–Sn chains as ScV$_6$Sn$_6$[34], provides an ideal platform for such investigations.

In this work, we combine angle-resolved photoemission spectroscopy (ARPES), scanning tunneling microscopy (STM), and density functional theory (DFT) calculations to study the CDW state of LuNb$_6$Sn$_6$. We find no charge instabilities associated with the vHSs near $E_F$, ruling out electronic interactions as the origin of the CDW. Moreover, we reveal that the band structure modifications in response to the emergence of CDW order have orbital selectivity, where the Lu 5*d* and Sn 5*p* electrons exhibit pronounced evolutions and the Nb 4*d* electrons remain unchanged. Following these observations, we suggest that the Nb kagome lattice is not relevant to the CDW in LuNb$_6$Sn$_6$, which is most likely driven by the structural instabilities of Lu and Sn atoms. Our results point to a universal CDW mechanism in all the charge-ordered "166" kagome materials, providing valuable insights into the interplay between electron, lattice, and orbital degrees of freedom in the CDW state of kagome metals.

**Results**

Figure 1a displays the pristine crystal structure of LuNb$_6$Sn$_6$, which is isostructural to ScV$_6$Sn$_6$ and belongs to the family of *P*6/*mmm* HfFe$_6$Ge$_6$-type[52] "166" kagome metals. As shown in Fig. 1a(i), LuNb$_6$Sn$_6$ consists of an alternative stacking of LuSn$_2$, kagome Nb$_3$Sn, and hexagonal Sn layers along the *c* axis. Analogous to other "166" compounds, like GdV$_6$Sn$_6$[53] and ScV$_6$Sn$_6$[41,42], we find that three possible surface terminations can be obtained upon cleaving LuNb$_6$Sn$_6$ crystals, denoted as NbSn, LuSn, and Sn terminations, as illustrated in Fig. 1a(ii). To characterize the underlying CDW transition, we conducted heat capacity measurements on LuNb$_6$Sn$_6$. As depicted in Fig. 1b, an anomaly is revealed at ~55 K, a similar temperature comparable to that in the previous report[51], implying the onset of CDW formation. To visualize the in-plane superlattice modulation of the CDW (Fig. 1c), we further carried out the low-temperature STM measurements on LuNb$_6$Sn$_6$. As shown in Fig. 1d, the topographic image on the Sn layer identifies the honeycomb lattice with individual Sn atoms clearly resolved. The corresponding Fourier transform (inset of Fig. 1d) exhibits a set of CDW peaks (red circles) located at (1/3, 1/3) with respect to the Bragg peaks (green circles), corroborating the presence of a √3 × √3 *R*30° reconstruction in the CDW phase, similar to the case of ScV$_6$Sn$_6$[34,36,37].

To investigate in more detail the electronic properties of LuNb$_6$Sn$_6$, we performed the synchrotron-based ARPES measurements. The small beam spot of our ARPES experiments allows us to recognize the three types of terminations of LuNb$_6$Sn$_6$, which can be distinguished by the line shape of Sn 4*d* core levels[42,53]. The experimentally revealed electronic structures from two different terminations (NbSn and LuSn) of LuNb$_6$Sn$_6$ are presented in Fig. 2 (see Supplementary Fig. S1 for the results of Sn termination). By comparing the Fermi surface (FS) mappings from NbSn [Fig. 2a(iii)] and LuSn [Fig. 2b(iii)] terminations, one can see that, although the FSs from LuSn termination exhibit richer topologies compared to NbSn termination, the characteristic band feature of a kagome lattice manifested as the corner-sharing triangular pockets centered at zone corner is clearly observed on both terminations. In Figs. 2a(iv) and 2b(iv), we show the corresponding overall band dispersions along the $\bar{\Gamma}$-$\bar{K}$-$\bar{M}$-$\bar{\Gamma}$ directions. Similar to other "166" materials[37,42,53,54], the two terminations exhibit dramatically different valence band structures. However, we could still identify similar characteristic kagome bands from both terminations, including the DP1 (at about -0.05 eV) and DP2 (at about -0.5 eV) at $\bar{K}$ point and the vHS1 slightly above $E_F$ at $\bar{M}$ point. In general, these ARPES spectra show a reasonably good agreement with our DFT calculations, as seen in Fig. 2f.

The most prominent differences between the ARPES mappings from NbSn [Fig. 2a(iii)] and LuSn [Fig. 2b(iii)] terminations are the existence of two FS contours around the second BZ center in the latter. For a better visualization of these pockets, we present the constant-energy intensity plots collected at $h\nu$ = 50 eV. As shown in Fig. 2c, a circular FS sheet and a hexagonal FS sheet near the second $\bar{\Gamma}$ point are observed (as labelled by the blue arrows), which arise out of the two electron-like bands as indicated by the orange dashed curve in Fig. 2e. These features resemble the nontrivial $Z_2$ topological surface states (TSSs) in GdV$_6$Sn$_6$[53] and ScV$_6$Sn$_6$[37]. The surface state nature can be verified by the photon-energy dependent ARPES measurements. In Fig. 2d, we present the FS mapping in the $k_z$-$k_{//}$ plane with the $k_{//}$ oriented along the $\bar{M}$-$\bar{\Gamma}$-$\bar{M}$ direction. One sees that the electron-like surface bands do not disperse with the increase of photon energy (as guided by the orange dashed lines in Fig. 2d), in contrast to the bulk states around $\bar{M}$ point (green dashed curve in Fig. 2d). To study the electronic properties of CDW ordering, below we focus on the bulk band structure of LuNb$_6$Sn$_6$.

The formation of 2 × 2 × 2 CDW order in the kagome metals $A$V$_3$Sb$_5$ and FeGe has been proposed to be intimately associated with the electronic instabilities arising from the nesting of multiple vHSs near $E_F$[10,11,29,32,33]. In contrast, prior studies of ScV$_6$Sn$_6$ have suggested the essential role of the lattice degrees of freedom other than the interactions between vHSs in driving the charge order therein[42–45]. To examine the role of vHSs in LuNb$_6$Sn$_6$, we first analyzed the low-energy electronic structures near the BZ boundary to understand the saddle-point topology of the vHSs. In Fig. 3a, as illustrated by a three-dimensional stack of the ARPES spectra taken across $\bar{M}$ point (parallel to the $\bar{\Gamma}$-$\bar{M}$-$\bar{\Gamma}$ direction, indicated by cuts #1-#5 in the inset of Fig. 3d), a series of hole-like dispersions (orange dashed curves) with the band tops showing the minimum energy at about 0.05 eV at $\bar{M}$ point are identified, tracing out an electron-like dispersion (orange solid curve) along the $\bar{K}$-$\bar{M}$-$\bar{K}$ direction. These bands thus form a saddle point, i.e., the vHS1, at $\bar{M}$ point. For comparison, in Figs. 3b and 3c, we also carried out the similar analysis on ScV$_6$Sn$_6$, where two vHSs are recognized below $E_F$ (both at about -0.04 eV), consistent with previous reports[37,42]. In order to study whether these vHSs close to $E_F$ can induce any electronic instabilities in LuNb$_6$Sn$_6$ and ScV$_6$Sn$_6$, we calculated the zero-frequency joint density of states (DOS) by the autocorrelation of the experimental FSs:

$$C(\boldsymbol{q}, E_F) = \int A(\boldsymbol{k}, E_F) A(\boldsymbol{k} + \boldsymbol{q}, E_F) d\boldsymbol{k}, \qquad (1)$$

where $A(\mathbf{k}, E_F)$ is the spectral function at $E_F$ at the $\mathbf{k}$ point in the BZ. The zero-frequency joint DOS describes the phase space for scattering of electrons from the state at $\mathbf{k}$ to the state at $\mathbf{k} + \mathbf{q}$ by certain modes with wave vector $\mathbf{q}$. Therefore, one can expect that $C(\mathbf{q}, E_F)$ peaks at the corresponding ordering wave vector if there exists an electronic instability due to the FS nesting. This autocorrelation of the ARPES spectra has been demonstrated to give a reasonable count for the charge-ordering instabilities of various compounds, like $AV_3Sb_5$[15], $FeGe$[32], cuprates[55,56], and transition metal dichalcogenides[57,58]. In Figs. 3d and 3e, we show the autocorrelation maps of $LuNb_6Sn_6$ and $ScV_6Sn_6$, respectively. It is revealed that, apart from the in-site local coherence peak at $\mathbf{q} = 0$, there are no additional peaks throughout the BZ for both compounds, indicating the absence of charge instabilities induced by the nesting between the vHSs near $E_F$. Although possessing different vHS positions, neither $ScV_6Sn_6$ nor $LuNb_6Sn_6$ shows any $C(\mathbf{q}, E_F)$ peaks from FS nesting according to the joint DOS results, even in the presence of more delocalized $4d$ Nb electrons, thus ruling out charge instability as the driving mechanism for the CDW ordering in these two compounds.

We now examine the intricate effect of CDW order on the electronic structures of $LuNb_6Sn_6$ by performing the temperature-dependent ARPES measurements across the CDW transition. Figures 4a(i) and 4a(ii) display the high-resolution ARPES spectra and its second derivative of $LuNb_6Sn_6$ in the pristine phase along the $\bar{K}$-$\bar{M}$-$\bar{K}$ direction ($T$ = 90 K, $h\nu$ = 67 eV), respectively. Due to the $k_z$ broadening effect, particularly in the vacuum ultraviolet regime[59], the ARPES spectra actually reflect the electronic states integrated over a certain $k_z$ region of the bulk BZ[60], with the electronic states at $k_z = 0$ and $\pi$ making the main contributions. In this context, we overlay the orbital-resolved DFT calculations for $k_z = 0$ and $\pi$ planes in Fig. 4d. One sees that, most experimental features around $\bar{M}$ point (Fig. 4a) can be well reproduced by the DFT results of $k_z = 0$ (Fig. 4d), including the $\alpha$, $\varepsilon$ bands crossing $E_F$ and the $\gamma$, $\kappa$ bands away from $E_F$, except for the $\beta$ band which is the projection from $k_z = \pi$. Upon entering the CDW phase [Figs. 4b(i) and 4b(ii), $T$ = 25 K], analogous to the case of $ScV_6Sn_6$[61], although no evident band reconstruction is observed along the $\bar{K}$-$\bar{M}$-$\bar{K}$ direction, the energy separation between the $\alpha$ and $\beta$ bands appears to be enlarged compared to that in the pristine phase. To validate and quantify the temperature evolution, as indicated by the red solid line in Fig. 4a(ii), we plot the corresponding energy distribution curves (EDCs) at 90 and 25 K in Fig. 4c(i) (red and blue dots). We fit the EDCs by using three Gaussian peaks (green, blue, and pink solid curves) and a background (brown solid curve), which is modeled by considering a polynomial function together with the Fermi-Dirac distribution. The fitting results are superimposed as the black solid curves. As highlighted by the black dashed lines in Fig. 4c(i), the $\alpha$ band shows minimal change across $T_{CDW}$; in contrast, a downward shift of ~20 meV of the $\beta$ band is clearly observed; moreover, the $\gamma$ band also exhibits a downward shift of ~12 meV. These evolutionary trends are further corroborated by the analysis of the momentum distribution curves (MDCs) presented in Fig. 4c(ii).

## Discussion

To better understand the electronic properties of CDW in $LuNb_6Sn_6$, we carried out the band unfolding calculations in the √3 × √3 × 3 CDW phase. As shown in Fig. 4e (see Supplementary Fig. S2 for the calculations along other high-symmetry lines), the DFT calculations well capture our experimental observations, reproducing the nearly intact α band ($k_z$ = 0) and the downshifted β ($k_z$ = π) and γ ($k_z$ = 0) bands across the CDW transition. The good correspondence between experiments and theory demonstrates that the revealed temperature evolutions are indeed the CDW-related features. By examining their orbital characters (Fig. 4d), we identify that the α band is solely derived from the Nb 4$d$ orbitals, while the β and γ bands contain not only the Nb 4$d$ orbitals but also non-negligible contributions from the Lu 5$d$ (β and γ) and Sn 5$p$ (γ) orbitals. These results suggest that the distinct band structure modifications (α and β/γ) in response to the CDW formation are most likely orbital selective, with the marginal change of the Nb 4$d$ electrons and the noticeable evolutions of the Lu 5$d$ and Sn 5$p$ electrons across $T_{CDW}$. The almost unaltered Nb kagome bands are also compatible with the negligible electronic interactions of the kagome vHSs, which, taken together, point to a minor role of the Nb kagome lattice in driving the CDW of LuNb$_6$Sn$_6$. Therefore, the revealed orbital selectivity of the CDW's impact can further signify the essential role of the structural components associated with the Lu and Sn sites in triggering the CDW transition. Consistently, the dominant structural distortions involving the Lu and Sn atoms other than the Nb atoms have been unveiled by the X-ray refined low-temperature structure of LuNb$_6$Sn$_6$[51]. Such findings are similar to the case of ScV$_6$Sn$_6$[34], evincing the identical structural origin of the charge orders in both compounds.

The CDW-induced band reconstruction and gap opening have been identified on the Sn 5$p$ bands around $\bar{\varGamma}$ point in ScV$_6$Sn$_6$[39,41,42]. Thus, it is also instructive to study whether these features can be observed in LuNb$_6$Sn$_6$. As illustrated by the unfolded and orbital-projected DFT calculations in Supplementary Figs. S2 and S3, a hole-like band η around $\varGamma$ point, which embraces a non-negligible contribution from the Sn 5$p$ orbitals, is proposed to harbor a charge order gap at around -0.5 eV in the CDW phase. Accordingly, in Figs. 4f(i) and 4f(ii), we show the ARPES intensity plots and their second derivatives along the $\bar{K}$-$\bar{\varGamma}$-$\bar{K}$ lines well above and below $T_{CDW}$, respectively. The η band and the neighboring electron band are clearly revealed, agreeing well with the DFT. However, the η band exhibits no visible band reconstruction or gap opening upon crossing $T_{CDW}$. This is in sharp contrast to the observation of CDW band gaps on the Sn 5$p$ bands in ScV$_6$Sn$_6$[39,41,42] (as also seen in Supplementary Fig. S4). In general, the energy gaps tied to the charge modulations are caused by the hybridizations between the original bands and the CDW folded bands, with the resulting gap size being determined by the underlying EPC strength[62,63]. In this context, the absence of CDW gap opening unveiled here suggests that the CDW modulation potential and EPC in LuNb$_6$Sn$_6$ are most likely much weaker than those in ScV$_6$Sn$_6$, although they share a similar CDW mechanism.

In summary, we have systematically studied the electronic structure of charge-ordered kagome metal LuNb$_6$Sn$_6$ and its evolution across the CDW transition. The comprehensive characterization of the electronic properties unequivocally uncovers that the primary driving force behind the charge order is the structural components similar as in ScV$_6$Sn$_6$, distinct from the CDW in $A$V$_3$Sb$_5$ and FeGe, where the electronic interactions of kagome electrons play an essential role. Moreover, we unambiguously reveal the orbital-selective band structure modifications across $T_{CDW}$, with the Lu 5$d$ and Sn 5$p$ electrons undergoing prominent evolutions while the Nb 4$d$ electrons remaining intact, underscoring the dominant contributions of Lu and Sn atomic displacements to the structural instabilities. Our results establish a universal origin of the underlying charge ordering in a broad class of "166" kagome systems, inspiring further exploration of emergent quantum phenomena in kagome lattice materials with structural instabilities and strong EPC.

The similarity between LuNb$_6$Sn$_6$ and ScV$_6$Sn$_6$ is reminiscent of another two isostructural "135" kagome lattice systems, $A$V$_3$Sb$_5$ and $A$Ti$_3$Bi$_5$. In stark contrast to $A$V$_3$Sb$_5$, whose CDW order arises primarily from the nesting of vHSs near $E_F$[10,11,33], the vHSs in $A$Ti$_3$Bi$_5$ are reported to reside well away from $E_F$, and thus no electronic instabilities and charge ordering have been observed therein[64,65]. However, regarding the "166" family, there exist analogous CDW wave vector and CDW mechanism in LuNb$_6$Sn$_6$ and ScV$_6$Sn$_6$, most likely also in other potentially charge-ordered compounds, despite very different unit-cell volumes[34,51]. The difference in lattice parameters is expected to result in distinct phonon dispersions, which may alter the properties of CDW, but here in the "166" compounds, the unstable phonon mode associated with the Lu(Sc)–Sn $c$-axis distortions appears to be less sensitive to the extrinsic parameters like chemical substitution or pressure. It can therefore be deduced that the nesting driven CDW state, the formation of which relies on multiple factors such as FS nesting conditions, electronic dimensionality, and EPC strength, seems to be more susceptible to the perturbations than the structurally driven CDW state revealed here. In this context, exploring the entanglement between robust charge ordering and other emergent states, realized through further engineering of the "166" systems, could open a new avenue for tailoring exotic physical properties.

## Methods

### Sample synthesis and characterization

LuNb$_6$Sn$_6$ single crystals were grown using the self-flux method as detailed in Ref. 51. Small pieces of Lu chunks (Alfa, 99.9%) were combined with Nb powder (Alfa, 99.9%) and Sn shots (Alfa, 99.999%) in a molar ratio of Lu:Nb:Sn = 8:2:90. The mixture was placed into an alumina crucible, which was then sealed in a vacuum quartz ampoule. The ampoule was heated to 1150 °C over 10 hours and held at this temperature for 24 hours. It was then cooled to 900 °C at a rate of 0.8 °C/h. The excess Sn flux was removed using a centrifuge. Heat capacity measurements were conducted using the HC option of a Physical Property Measurement System (PPMS, Quantum Design). Single crystals of ScV$_6$Sn$_6$ were also grown via the flux-based growth technique, as described in Ref. 41.

### ARPES measurements

High-resolution ARPES measurements of LuNb$_6$Sn$_6$ were performed at the BLOCH beamline of MAX IV Laboratory and the 1$^2$- and 1$^3$-ARPES end stations of UE-112-PGM2 beamline at Helmholtz Zentrum Berlin BESSY-II light source. High-resolution ARPES measurements of ScV$_6$Sn$_6$ were conducted at the BL03U beamline of the Shanghai Synchrotron Radiation Facility. The energy and angular resolutions were set to better than 5 meV and 0.1°, respectively. Samples were cleaved *in* *situ*, yielding flat mirrorlike (001) surfaces. During the experiments, the temperature was kept at 18-25 K if not specified otherwise, and the vacuum conditions were maintained better than 7.0 × 10$^{-11}$ Torr. We used linear horizontal polarization for all the measurements.

**STM measurements**

Low-temperature STM measurements were performed using a Unisoku USM-1300 system, with a base pressure of 1.0 × 10$^{-10}$ Torr. The samples were cleaved mechanically *in situ* at room temperature, and then immediately inserted into the STM head. Topographic images were obtained with Pt/Ir tips.

**Band structure calculations**

The first-principles calculations were performed based on the DFT as implemented in the Vienna ab initio simulation package (VASP) using the projector augmented wave (PAW) method[66–68]. The generalized gradient approximation (GGA) with the Perdew-Burke-Ernzerhof (PBE) scheme was adopted for the exchange-correlation functional[69]. The cutoff energy was set to be 450 eV with $\Gamma$-centered $k$ mesh. The energy and force convergence criteria were set to be 10$^{-6}$ eV and 10$^{-2}$ eV/Å, respectively. All the results were obtained after the structure relaxation of experimental structure as provided in Ref. 51.

**Data availability**

All data needed to evaluate the conclusions in the paper are present in the paper and the Supplementary Information file. All raw data generated during the current study are available from the corresponding authors upon request.

**Code availability**

The computer codes used for the band structure calculations in this study are available from the corresponding authors upon request.

**Acknowledgements**

We acknowledge MAX IV Laboratory for the experimental time on BLOCH beamline under proposal 20240299. Research conducted at MAX IV, a Swedish national user facility, was supported by the Swedish Research Council under Contract No. 2018-07152, the Swedish Governmental Agency for Innovation Systems under Contract No. 2018-04969, and Formas under Contract No. 2019-02496.

The authors are particularly grateful for the expert scientific and technical support of Craig Polley and the entire BLOCH team. This work was supported by the Deutsche Forschungsgemeinschaft under Grant SFB 1143 (project C04) and the Würzburg-Dresden Cluster of Excellence on Complexity and Topology in Quantum Matter − *ct.qmat* (EXC 2147, project ID 390858490). E.J.C. and X.L.F. acknowledge the financial support from the Alexander von Humboldt Foundation. S.B. and B.B. acknowledge the support from the BMBF via project UKRATOP. Z.K.Liu acknowledges the support from the National Natural Science Foundation of China (Grants No. 92365204 and No. 12274298) and the National Key R&D program of China (Grant No. 2022YFA1604400/03).

## Author contributions

R.L., E.J.C., and Z.K.Liu conceived the project. E.J.C., Y.X.L., Q.X.W., Y.Z., W.S., R.K., D.C., Y.F.G., and C.F. synthesized the single crystals and performed the transport measurements. R.L., Y.M.Z., A.F., A.G., S.C., Y.L.C., S.B., D.V.V., B.B., and Z.K.Liu carried out the ARPES experiments. R.L. and Y.M.Z. analyzed the ARPES data. Z.K.Li conducted the STM measurements. X.L.F., H.Y.M., and J.P.L. performed the DFT calculations. R.L., E.J.C., X.L.F., and Z.K.Liu supervised the project. R.L. wrote the paper with input from all authors.

## Competing interests

The authors declare no competing interests.

## Additional information

**Supplementary information** is available in the online version of the paper.

**Correspondence** and requests for materials should be addressed to Rui Lou, Erjian Cheng, Xiaolong Feng or Zhongkai Liu.

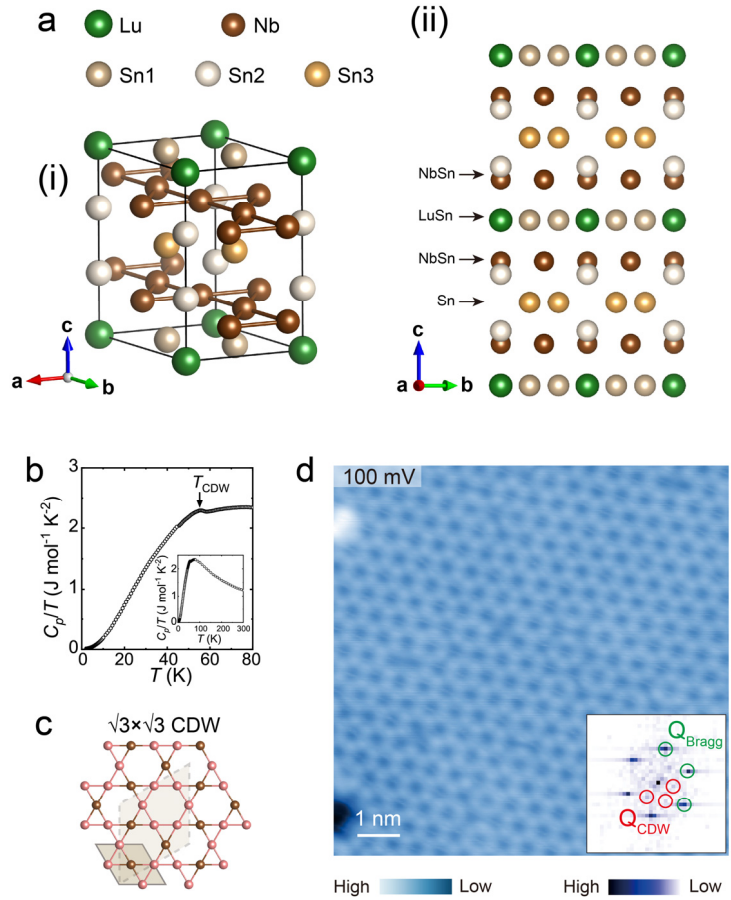

**Figure 1 | Crystal structure and CDW phase of LuNb$_6$Sn$_6$. a,** Crystal structure of LuNb$_6$Sn$_6$ (i) and its side view along the [100] direction (ii). Three types of Sn atoms with different chemical environments are marked out, namely, the Sn1 atom in LuSn$_2$ layer, the Sn2 atom in Nb$_3$Sn layer, and the Sn3 atom in Sn layer. **b,** Zero-field specific heat of LuNb$_6$Sn$_6$ in a form of $C_p/T$ versus $T$. The onset of CDW transition at ~55 K is revealed. Inset shows the same data over a wide temperature range. **c,** Schematic illustration of the in-plane charge ordering of LuNb$_6$Sn$_6$. The brown and grey shades indicate the pristine and $\sqrt{3} \times \sqrt{3}$ $R30°$ unit cells, respectively. **d,** Atomically resolved STM topography on the Sn termination taken at $T$ = 4.2 K ($V_s$ = 100 mV, $I$ = 1 nA). Inset shows the corresponding Fourier transform with the CDW and atomic Bragg peaks highlighted by the red and green circles, respectively.

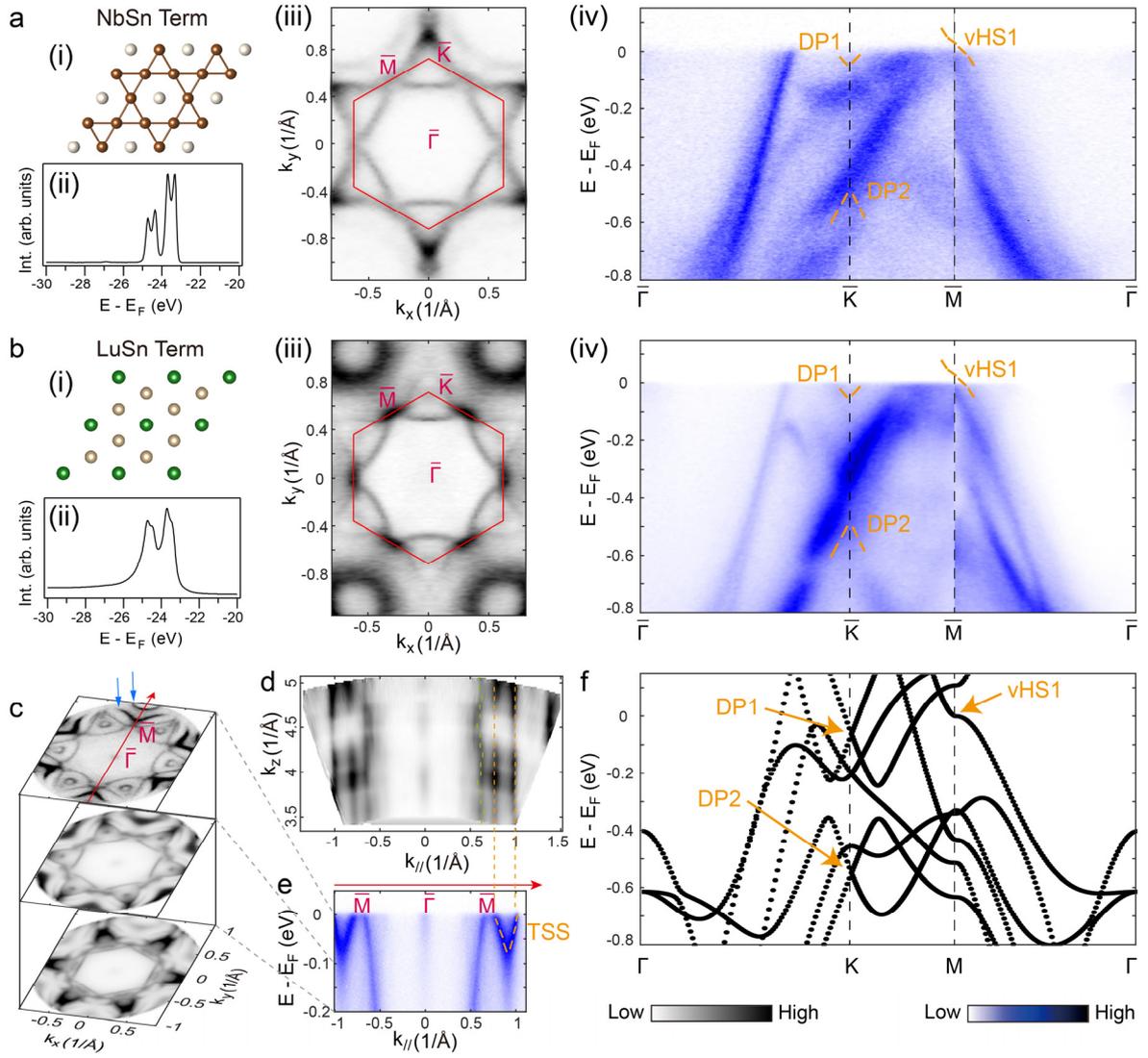

**Figure 2 | Overall band structure from different terminations of LuNb$_6$Sn$_6$. a, b,** Typical electronic properties obtained on the NbSn and LuSn terminations, respectively. (i), (ii) Schematics of the two terminations and corresponding Sn 4$d$ core-level spectra, respectively. (iii) FS mappings recorded by 67- (NbSn) and 80-eV (LuSn) photons. The red solid curves indicate the pristine BZs. (iv) A summary of the electronic structures of LuNb$_6$Sn$_6$ along the $\bar{\Gamma}$-$\bar{K}$-$\bar{M}$-$\bar{\Gamma}$ lines at $T$ = 20 K. As guided by the orange dashed curves, the typical kagome bands (DP1, DP2, and vHS1) are observed on both terminations. **c,** Stack of the constant-energy ARPES contours ($h\nu$ = 50 eV) from the LuSn termination at the energies of 0, -0.1, and -0.2 eV, respectively. The red solid arrow indicates the in-plane momentum orientation of the $k_z$ mapping in **d**. **d,** ARPES intensity plot in the $k_z$-$k_{//}$ plane at $E_F$, with $k_{//}$ oriented along the $\bar{M}$-$\bar{\Gamma}$-$\bar{M}$ direction. The orange and green dashed curves are guides to the eye for the TSS around $\bar{\Gamma}$ and a bulk state around $\bar{M}$, respectively. The inner potential $V_0$ is estimated to be about 12 eV. **e,** ARPES spectra taken along the $\bar{M}$-$\bar{\Gamma}$-$\bar{M}$ line with 50-eV photons. The TSS is traced out by the orange dashed curve. **f,** DFT calculated bulk band structure of LuNb$_6$Sn$_6$ in the pristine phase along the $\Gamma$-$K$-$M$-$\Gamma$ lines. As highlighted by the orange arrows, the DP1, DP2, and vHS1 revealed in **a**(iv) and **b**(iv) are well reproduced by the calculations.

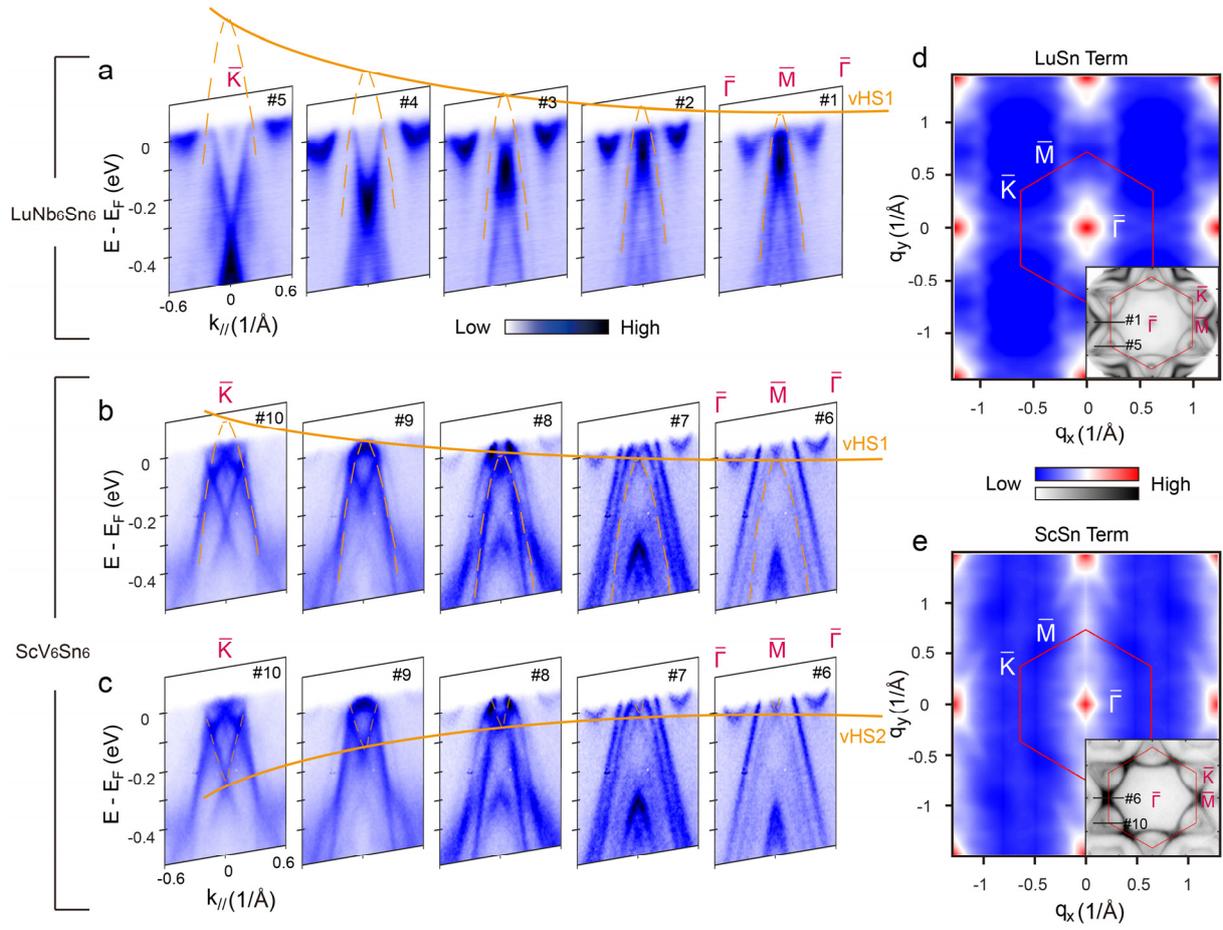

**Figure 3 | Characterization of the near-$E_F$ vHSs in LuNb$_6$Sn$_6$ and ScV$_6$Sn$_6$. a,** Stack of ARPES intensity plots ($h\nu$ = 50 eV) of LuNb$_6$Sn$_6$ taken perpendicular to the $\bar{K}$-$\bar{M}$-$\bar{K}$ direction across the BZ boundary, as indicated by the evenly spaced cuts #1-#5 in the inset of **d** with the cut #1 passing through $\bar{M}$ point and the cut #5 passing through $\bar{K}$ point. The orange solid and dashed curves mark out the dispersions of the vHS1 along and perpendicular to the $\bar{K}$-$\bar{M}$-$\bar{K}$ lines, respectively. **b,** Dispersions across a m-type vHS of ScV$_6$Sn$_6$. The spectra are measured by 82-eV photons along the evenly spaced cuts #6-#10 in the inset of **e**, with the cut #6 crossing $\bar{M}$ point and the cut #10 crossing $\bar{K}$ point. The orange solid and dashed curves are guides to eye for the dispersions of the vHS along and perpendicular to the $\bar{K}$-$\bar{M}$-$\bar{K}$ paths, respectively. **c,** Same data as in **b** identifying a p-type vHS of ScV$_6$Sn$_6$. **d,** Two-dimensional joint DOS results from the experimental FS of LuNb$_6$Sn$_6$ with the LuSn termination. Inset shows the corresponding FS mapping data. **e,** Same as **d** of ScV$_6$Sn$_6$ with the ScSn termination.

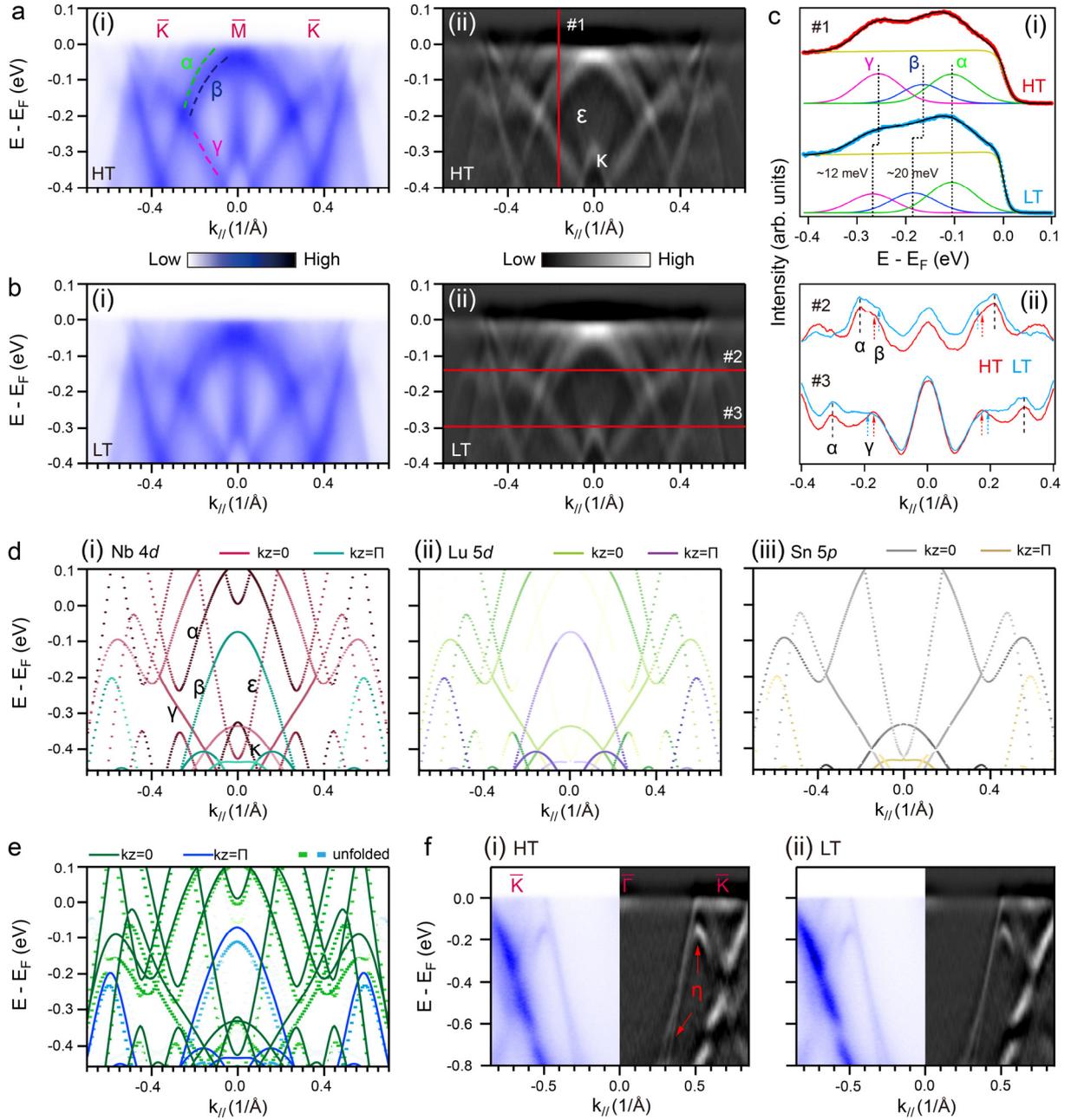

**Figure 4 | CDW-induced band structure modifications in LuNb$_6$Sn$_6$. a, b,** ARPES intensity plots (i) and corresponding second derivative intensity plots (ii) of LuNb$_6$Sn$_6$ ($h\nu$ = 67 eV, LuSn termination) along the $\overline{K}$-$\overline{M}$-$\overline{K}$ directions measured at $T$ = 90 and 25 K, respectively. The dashed curves are guides to the eye for the α, β, and γ bands. The red solid lines indicate the locations where the EDCs (#1) and MDCs (#2 and #3) in **c** are extracted. **c,** Temperature dependent EDCs (i) and MDCs (ii) taken at the red solid lines in **a**(ii) and **b**(ii). The EDCs are quantitatively fitted by using three Gaussian peaks (green, blue, and pink curves) and a background (brown curve). The black dashed lines in (i) indicate the evolution of peak positions. In (ii), the black dashed lines show the unshifted MDC peaks of α band; the red and blue dashed arrows mark out the MDC peak positions of β/γ band above and below $T_{CDW}$, respectively. **d,** DFT electronic structures of pristine LuNb$_6$Sn$_6$ along the *K-M-K* and *H-L-H* directions with spectral weight projected onto the Nb 4*d* (i), Lu 5*d* (ii), and Sn 5*p* (iii) orbitals, respectively. **e,** DFT calculated bands of LuNb$_6$Sn$_6$ along the *K-M-K* and *H-L-H* lines in the pristine

(solid curves) and CDW (dots) phases. The band structure in the CDW state is unfolded to the pristine BZ. **f,** ARPES spectra (negative momentum sides) and second-derivative spectra (positive momentum sides) of LuNb$_6$Sn$_6$ recorded along the $\overline{K}$-$\overline{\Gamma}$-$\overline{K}$ directions at $T$ = 90 (i) and 25 K (ii) under 50-eV photons, respectively.